\documentclass[twocolumn]{aastex63}

\shorttitle{Formation time}
\shortauthors{Spalding \& Winn}

\usepackage{hyperref}
\usepackage{comment}
\usepackage{graphicx}
\usepackage[titletoc]{appendix}

\usepackage{amsmath}
 
\begin{document}

\title{Tidal erasure of stellar obliquities constrains the timing of hot Jupiter formation} 

\author{Christopher Spalding} 
\author[0000-0002-4265-047X]{Joshua N.\ Winn}
\affil{Department of Astrophysical Sciences\\
Princeton University, Princeton, NJ 08540} 
\begin{abstract}
Stars with hot Jupiters sometimes have high obliquities, which are possible relics of hot Jupiter formation. Based on the characteristics of systems with and without high obliquities, it is suspected that obliquities are tidally damped when the star has a thick convective envelope, as is the case for main-sequence stars cooler than $\sim$6100K, and the orbit is within $\sim$8 stellar radii. A promising theory for tidal obliquity damping is the dissipation of inertial waves within the star's convective envelope. Here, we consider the implications of this theory for the timing of hot Jupiter formation. Specifically, hot stars that currently lack a convective envelope possess one during their pre-main sequence. We find that hot Jupiters orbiting within a critical distance of $\sim$0.02~au from a misaligned main-sequence star lacking a thick convective envelope must have acquired their tight orbits after a few tens of millions of years in order to have retained their obliquities throughout the pre-main-sequence. There are 4 known systems for which this argument applies--XO-3b, Corot-3b, WASP-14b, and WASP-121b--subject to uncertainties surrounding inertial wave dissipation. Moreover, we conclude that a recently-identified overabundance of near-polar hot Jupiters is unlikely sculpted by tides, instead reflecting their primordial configuration. Finally, hot Jupiters arriving around cool stars after a few 100s of millions of years likely find the host star rotating too slowly for efficient obliquity damping. We predict that the critical effective temperature separating aligned and misaligned stars should vary with metallicity, from 6300\,K to 6000\,K as [Fe/H] varies from $-0.3$ to $+0.3$.
\\ 
\end{abstract}

\section{Introduction}

The origin of hot Jupiters -- gas giants with orbital periods shorter than about 10 days -- has been debated since the beginning of exoplanetary science \citep{rasio1996dynamical,mayor1995jupiter,dawson2018origins}. Hot Jupiters are typically thought to have been constructed at larger distances, beyond 1~au, where an ample supply of solids and gas facilitates runaway core accretion \citep{pollack1996formation}. Afterward, these planets are thought to have moved inwards to arrive on their tight orbits. One of the key questions in the debate over hot Jupiter formation is \textit{when} this inward ``migration'' took place \citep{dawson2018origins}.

Hot Jupiter migration theories differ in the timing of formation. In ``high-eccentricity migration,'' a distant giant planet is dynamically driven onto a highly eccentric orbit, which then shrinks and circularizes due to tidal dissipation within the planet \citep{rasio1996dynamical,wu2003planet,fabrycky2007shrinking}. This can occur over a wide range of timescales, from millions to billions of years, depending on the system.
Alternatively, in ``disk-driven migration,'' hot Jupiters spiral inwards as an outcome of gravitational interactions with the protoplanetary disk  \citep{goldreich1980disk,lin1996orbital,baruteau2014planet}. Naturally, disk-driven migration must occur within a few million years of the star's formation when the disk is still present \citep{haisch2001disk,mamajek2009initial}.
A third option is that the giant planet forms in a short-period orbit, despite the initial theoretical assumption that this cannot occur \citep{batygin2016situ,boley2016situ}. In this scenario, too, the formation takes place while the gaseous disk still exists.

Even after decades of scrutiny, we do not know how frequently (if at all) each of these scenarios plays out \citep{dawson2018origins}. In this paper, we focus on an observational finding that might be brought to bear on this mystery: the orbital planes of hot Jupiters and the equatorial planes of their host stars are sometimes grossly misaligned \citep{hebrard2008misaligned,winn2009spin,winn2010hot,triaud2010spin}. This phenomenon, referred to as high stellar obliquities, orbital tilts, or spin-orbit misalignments,
may be a clue about the dynamical histories of these systems. 
Various mechanisms have been hypothesized to generate these obliquities, including primordial misalignments between the protoplanetary disk and the stellar equator \citep{batygin2012primordial,spalding2014early,lai2014star,fielding2015turbulent}, and inclination excitation that accompanies high-eccentricity migration \citep{dawson2014tidal,anderson2016formation,storch2017dynamics}. As with migration, the timescale of obliquity excitation is unknown and the theoretical expectations span a wide range.
Thus, the mere existence of large stellar obliquities does not rule out any particular planet formation pathway.

Among stars hosting hot Jupiters, those with effective temperatures hotter than about 6100\,K exhibit a broad range of obliquities; the angle between the sky projections of the spin and orbital axes
spans the full range from 0$^\circ$ to 180$^\circ$ \citep{albrecht2012obliquities}, and the true
obliquity ranges up to about 120$^\circ$ \citep{albrecht2021preponderance}.
In contrast, cooler stars typically have obliquities lower
than $30^\circ$ whenever the planet's orbit is smaller than about 8
stellar radii \citep{winn2010hot,albrecht2012obliquities,winn2015occurrence,Winn2017constraints,Louden+2021}.

\begin{figure*}[ht!]
\centering
\includegraphics[trim=0cm 0cm 0cm 0cm, clip=true,width=0.8\textwidth]{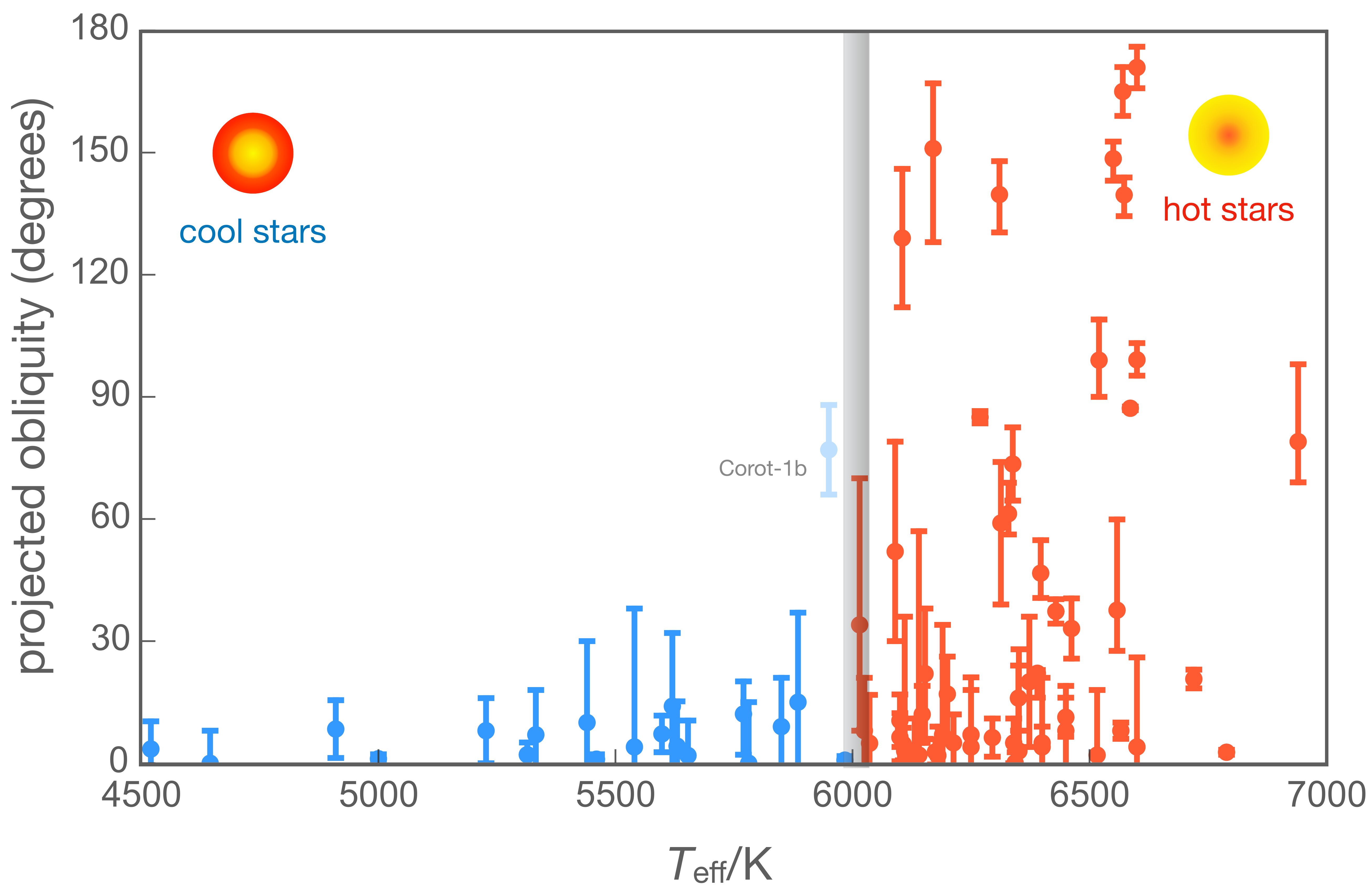}
\caption{Sky-projected stellar obliquity as a function of stellar effective temperature. The data shown are for all systems in the
TEPCAT database \citep{southworth2011homogeneous} with $m_p>0.3\,M_J$, $a_p<8R_\star$, and error bars smaller than $40^\circ$. Stars with
$T_{\rm eff}\lesssim 6100$K (vertical gray line) are typically aligned, while hotter stars exhibit a larger spread of obliquities. The faint blue point just below $6000$K is Corot-1b, for which one dataset suggests good alignment \citep{bouchy2008radial} and another indicates a strong misalignment \citep{pont2010spin}, and for which published determinations of the effective temperature differ by 300K \citep{barge2008transiting,torres2012improved,bonomo2017gaps}. For these reasons, we did not consider Corot-1 in our computations.}
\label{fig: Data1}
\end{figure*}

The data are shown in Figure~\ref{fig: Data1}. The transition temperature of 6100\,K, sometimes called the ``Kraft break" (after \citealt{kraft1967studies}), is also 
significant in stellar structure theory. The cooler stars have thick convective envelopes while the hotter stars have very thin convective zones. This internal difference leads to observable differences, such as the slower rotation rates of cool stars that is thought to be
a consequence of magnetic braking \citep{matt2015mass}.
These facts have been interpreted as evidence for
tidal dissipation of the obliquities of stars with thick convective
envelopes and especially close-orbiting giant planets \citep{winn2010hot,albrecht2012obliquities}.

A promising mechanism for tidal damping of obliquities was presented by \citet{lai2012tidal}. He found that a particular tidal oscillation mode --- the $\omega_{10}$ mode, in notation to be explained below --- acts
to damp the obliquity while avoiding catastrophic orbital decay,
a problem that had afflicted other tidal mechanisms.
Additionally, this mode oscillates with a frequency equal to the star's spin frequency, allowing it to launch
inertial waves\footnote{Inertial waves are disturbances for which the restoring force is the Coriolis force; see, e.g., \cite{ogilvie2009tidal}.}
within the star's convective zone. Dissipation of these waves eventually erases any initially high stellar obliquity \citep{ogilvie2013tides,lin2017tidal}
While the work of \citet{lai2012tidal} was mainly analytic,
subsequent numerical, statistical, and population-synthesis calculations have also been undertaken to study the possibility of tidal obliquity damping \citep{ogilvie2009tidal,ogilvie2013tides,lin2017tidal,penev2018empirical,anderson2021possible}.

The work described here was motivated by the realization that stars hotter than 6100\,K did have thick convective envelopes for tens of millions of years, when they were on the pre-main-sequence
\citep{amard2019first,amard2020impact}. Does the same theory that successfully explains the evidence for tidal obliquity damping in cooler main-sequence stars also predict that significant damping should have occurred in the remote past of hotter stars? If so, then the observation of a high obliquity of a hot star would imply that the hot Jupiter must have arrived after the star lost its convective envelope. This, in turn, would rule out \textit{in situ} formation and disk-driven migration.

In addition, tidal dissipation driven by inertial waves increases roughly quadratically with stellar spin rate \citep{ogilvie2013tides}. Cool stars spin down by over an order of magnitude over time \citep{gallet2013improved,matt2015mass}, and thus are only able to tidally lose their obliquities within a limited window of time, corresponding to several hundred million years after disk dispersal. This places an upper limit on the age at which hot Jupiters can arrive and still realign with their host stars.

In this work, we explore these two timing constraints quantitatively, delineating the parameter space within which hot Jupiters orbiting hot stars would be expected to have realigned if they formed early, and within which hot Jupiters would not have had time to realign if they formed late.  We find that astrophysically relevant constraints can indeed be obtained, though they are subject to large uncertainty because the theory of tidal dissipation of inertial waves is not yet developed enough to give unique predictions.  We also examine the relationship between stellar metallicity and the critical value of the effective temperature characterizing the Kraft break, as a possible empirical test of the tidal erasure hypothesis.

The rest of the manuscript is structured as follows. In Section 2, we describe our modeling framework. In Section 3, we describe the time-evolution of stellar spin, and in Section 4 we provide bounds on the magnitude of tidal dissipation. In Section 5, we present simulations the time evolution of stellar obliquity for real systems, on a system-by-system basis. In Section 6, we discuss the results
and in Section 7, we summarize our conclusions.

\section{Tidal theory}

In this section, we quantify the rate of obliquity-removal in stars as a function of convective zone thickness, including during the pre-main sequence. We make use of previously developed theory and numerical simulations of inertial wave-driven tidal dissipation \citep{goodman2009dynamical,ogilvie2009tidal,ogilvie2013tides,lin2017tidal}. Consider a hot Jupiter of mass $m_p$, orbiting a star of mass $M_\star$ on a circular orbit of semi-major axis $a_p$ that is inclined by some angle $\theta$ relative to the star's spin axis. In spherical coordinates ($r,\,\psi,\,\phi$), the planet's time-dependent tidal potential to quadrupolar ($l=2$) order takes the form \citep{polfliet1990dynamic,lai2012tidal,ogilvie2014tidal}

\begin{align}\label{Eq: tidal potential}
    \Psi=\frac{Gm_p}{a_p}\sum_{m=0}^{2}\sum_{m'}A_{mm'}(\psi)\bigg(\frac{r}{a}\bigg)^2Y_{2}^m\big(\psi,\phi\big)\mathrm{e}^{-im'\Omega_{\textrm{K}}t}.
\end{align}
Here, $A_{mm'}$ is a constant that depends upon the orbital inclination, $t$ is time, $\Omega_{\textrm{K}}$ is the Keplerian orbital angular frequency, and $G$ is Newton's gravitational constant. In the expression above, we have ignored terms of higher order than $\mathcal{O}(r/a)^2$ owing to the smallness of the stellar radius relative to the orbital distance. 
The spherical harmonic function in the inertial frame has an angular dependence $Y_{2}^m(\psi,\phi)\propto \exp(im \phi)$, but in the rotating frame the azimuthal angle is defined as $\phi_r=\phi-\Omega_\star t$. Substituting this transformation into $Y_{2}^m(\psi,\phi)$, we may write the potential $\Psi\propto \exp(im\phi_r-i\omega_{mm'} t)$, where we have defined the tidal frequency

\begin{align}
    \omega_{mm'}\equiv m'\Omega_{\textrm{K}}-m\Omega_\star.
\end{align}
 In response to these tidal components, the star is deformed, altering its gravitational potential. This altered potential can itself be expanded as a sum of harmonics, with each harmonic lagging the corresponding forcing harmonic by some time interval. Tidal theory's primary objective is to solve for the perturbed body's induced potential, the lag time and magnitude of which is conveniently written in terms of some mode-dependent, complex Love number, denoted $k_m^2$ to quadrupolar order \citep{ogilvie2013tides}. 
 
 The simplest approximation, known as the \textit{equilibrium} tide, asserts that all modes possess equal lag times, and thus the body essentially reaches a hydrostatic shape in response to the instantaneous pertubing potential  \citep{zahn1977tidal}. While convenient, this assumption would not suffice to explain the alignment of cool stars because equilibrium tides damp semi-major axis and obliquity at similar rates \citep{lai2012tidal}. Corrections to the equilibrium tide are referred to as \textit{dynamical} tides, which consist of non-hydrostatic, wavelike responses to the tidal potential \citep{ogilvie2014tidal}. Dissipation of these waves lead to specific values of $k_{2}^m$, depending on their physical nature, which in this work we consider to take the form of inertial waves dissipating in the convective envelope of the star \citep{goodman2009dynamical,ogilvie2013tides,lin2017tidal}.
 
We focus on the case with $m=1$ and $m'=0$, leading to the tidal frequency 
 \begin{align}
     \omega_{10}=-\Omega_\star.
 \end{align}
Crucially, the dissipation of this specific mode does not affect the orbital semi-major axis. Only the obliquity is damped. Accordingly, we refer to this mode as the ``obliquity tide". Inertial waves can be excited if $|\omega_{mm'}|\leq 2\Omega_\star$, which is always the case for $\omega_{10}$ \citep{lai2012tidal}, thereby guaranteeing a dynamical contribution to the obliquity tide.

\begin{figure*}
\centering
\includegraphics[trim=0cm 0cm 0cm 0cm, clip=true,width=0.9\textwidth]{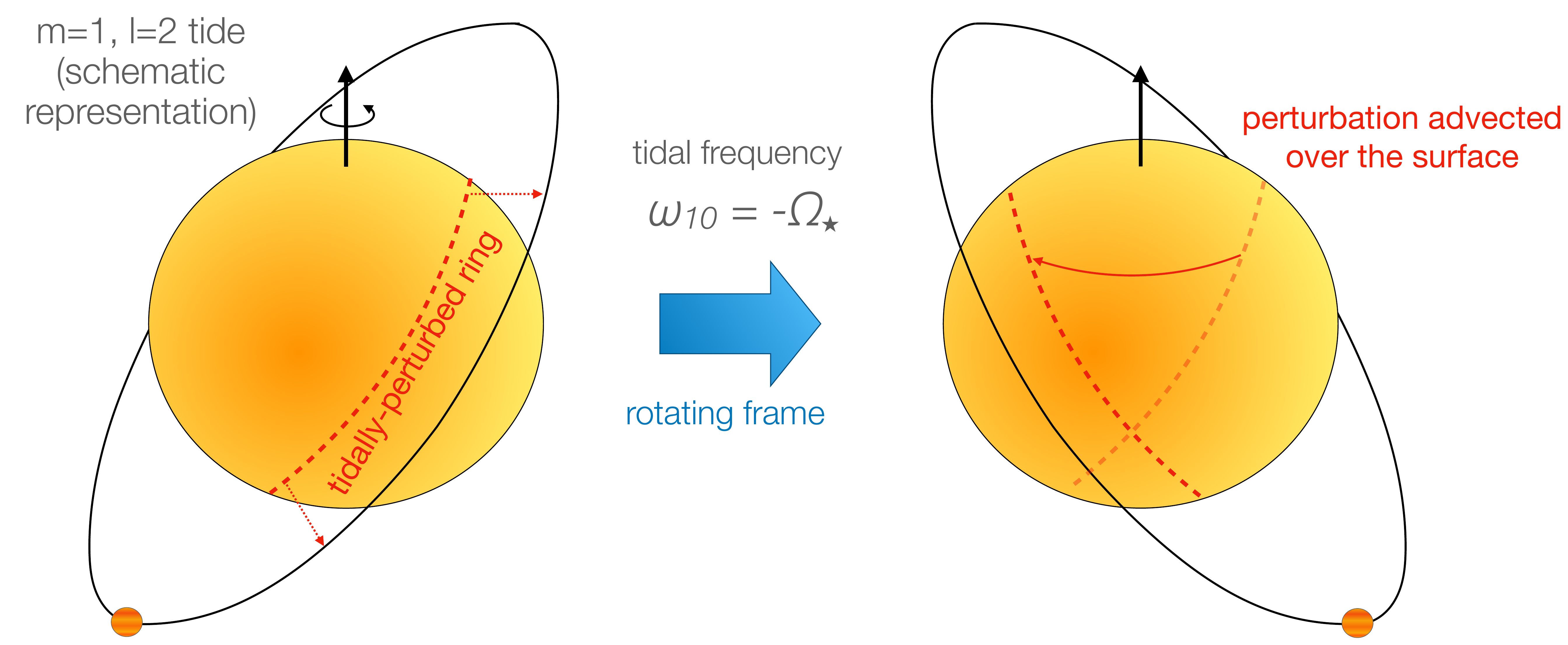}
\caption{Schematic representation of the $l=2$, $m=1$, $m'=0$ mode responsible for the obliquity tide. The mode is stationary in the inertial frame (left), and may be conceptualized as a tidal bulge that follows the great circle immediately below the planet's orbit. As the star rotates beneath the orbit, the bulge sweeps over the surface once per rotation period. Thus, in the fluid frame (right), the tidal frequency $\omega_{10}=-\Omega_\star$ which is within the range for inertial wave excitation.}
\label{fig: Schematic}
\end{figure*}

The physical origin of this mode is illustrated schematically in Figure~\ref{fig: Schematic}. In a reference frame rotating with the star, the planet's orbit-averaged potential is equivalent to that of a massive wire that precesses with frequency $-\Omega_\star$ about the stellar spin axis. This wire can be thought of as raising a bulge on the underlying stellar surface. Due to dissipation the response of the fluid surface lags behind the instantaneous position of the wire, generating a torque that damps the obliquity. The amplitude of the lag is determined by the physics of inertial wave launching, propagation, and dissipation in the convective zone \citep{goodman2009dynamical,ogilvie2013tides}.

To model the dissipation of this mode, we follow \citet{ogilvie2013tides} and note that the imaginary part of $k_{2}^m$ encodes the magnitude of tidal dissipation resulting from the tidal perturbation in Equation~\ref{Eq: tidal potential}. Written in terms of the more commonly-used hydrostatic degree-2 Love number $k_2$ and the tidal quality factor $Q_{mm}$, the potential Love number becomes \citep{ogilvie2014tidal}

\begin{align}\label{eq: def}
    \textrm{Im}\big[k_2^m\big]=\textrm{sign}(\omega_{mm'})\frac{k_2}{Q_{mm'}}.
\end{align}
 For brevity, henceforth we write Im$[k_2^m]$ simply as $k_2^m$.
 In response to the $m=1$, $m'=0$ perturbation identified above, the obliquity evolves according to the equation \citep{lai2012tidal}
 
\begin{align}\label{eq: obliquity}
    \dot{\theta}\big|_{10}&=-\frac{3}{4}\nu\frac{k_2}{Q_{10}}\sin \theta\cos^2\theta\bigg[1+\frac{L}{S}\cos\theta\bigg],
\end{align}
where $L\equiv m_p\sqrt{G M_\star a_p}$ and $S\equiv \mathcal{K}_\star M_\star R_\star^2\Omega_\star$ are the orbital and spin angular momenta. Here, $M_\star$ and $R_\star$ are the stellar mass and radius, and $\mathcal{K}_\star$ is the dimensionless moment of inertia constant. Moreover, we define the tidal rate as
\begin{align}
    \nu \equiv  \bigg(\frac{M_p}{M_\star}\bigg)\bigg(\frac{R_\star}{a}\bigg)^5\Omega_K.
\end{align}
The tidal rate and the stellar obliquity are related to changes in the stellar spin rate via \citep{lai2012tidal,ogilvie2014tidal}
\begin{align}\label{eq: spin-rate}
   \dot{\Omega}_\star\big|_{10}&=-\frac{3L}{4S}\Omega_\star\nu\frac{k_2}{Q_{10}}\big(\sin \theta\cos\theta\big)^2.
\end{align}
Together, equations~\ref{eq: obliquity} \&~\ref{eq: spin-rate} fully describe the evolution of the star in response to dissipation arising from the $\omega_{10}$ component of the obliquity tide.

\subsection{Dissipation from inertial waves}

In general, the value of $k_2^1$ is not computed analytically, but rather through numerical simulations based on a given model for the stellar interior \citep{ogilvie2013tides,lin2017tidal}. We consider a two-layer star possessing a convective zone atop a radiative core with radius $R_c$ and mass $M_c$ (we ignore any small convective core). We define the structural parameters $\alpha\equiv R_c/R_\star$ and $\beta\equiv M_c/M_\star$, which both tend to unity as the Kraft break is approached \citep{amard2019first}. 
The functional dependence of $k_2^1$ upon $\alpha$ and $\beta$ varies with the assumed stellar model and numerical technique \citep{ogilvie2013tides,lin2017tidal}, however, a robust result is that dissipation from inertial waves scales as $\epsilon^2$, where 

\begin{align}
    \epsilon\equiv \frac{\Omega_\star}{\sqrt{G M_\star/R_\star^3}}
\end{align}
is the ratio of the stellar spin rate to its break-up spin rate. 
Thus, we assume the magnitude of $k_2^1$ can be written \citep{ogilvie2013tides,mathis2015variation,lin2017tidal,damiani2018influence}

\begin{align}\label{eq: quadratic}
     k_{2}^1=\epsilon^2 \mathcal{F}_{21}(\alpha,\beta).
\end{align}
Unfortunately, $\mathcal{F}_{2 1}(\alpha,\beta)$ has not yet been computed for inertial wave dissipation and a realistic stellar structure model \citep{ogilvie2013tides,lin2017tidal,barker2020tidal}. Early, semi-analytic treatments of inertial wave-driven dissipation from the $l=2$, $m=2$ mode (as opposed to $m=1$) found that $k_{2}^2$ scales as the fifth power of the core radius \citep{goodman2009dynamical,ogilvie2009tidal}, but for analytical convenience these models were based
on an unrealistic (incompressible) model for the stellar envelope. 

More recently, numerical simulations have been undertaken to compute the magnitude of $k_2^1$ for convective shells within a frequency-averaged framework \citep{ogilvie2013tides,barker2020tidal}. \citet{ogilvie2013tides} considered the general, frequency-averaged response of barotropic fluids to a range of tidal harmonics, whereas \cite{barker2020tidal} computed the frequency-averaged response of realistic stellar interiors to the $m=2$ mode. These simulations did not reproduce the fifth-power scaling that had been found in the earlier semi-analytic calculations. However, the tidal response to $\omega_{10}$ is complicated by a resonance, referred to as the ``spin-over mode", which represents solid-body precession of the star and does not dissipate tidal energy \citep{lin2017tidal}. This resonance makes the frequency-averaged approach of \citet{ogilvie2013tides} inappropriate for our purposes.

A more specialized treatment of the $m=1$ mode associated with the obliquity tide, which is of more direct relevance to this work, was performed by \citet{lin2017tidal}. They removed the component of the tidal response associated with solid-body precession in order to compute the proportion of the fluid response that contributes to tidal evolution. Their procedure did recover the fifth-power dependence of dissipation rate upon core radius \citep{goodman2009dynamical,ogilvie2009tidal}. 

Given that the work of \citet{lin2017tidal} is the only available study that self-consistently accounts for the presence of the spin-over mode within the obliquity tide, we decided to assume that dissipation scales as the fifth power of core size (as described further below). Nevertheless, we acknowledge that a priority for future work should be numerical investigations of dissipation within realistic stellar interiors, as done by \citet{barker2020tidal}, while simultaneously accounting for the influence of the spin-over mode, following \citet{lin2017tidal}.

The fifth power scaling with core radius must break down as $R_c\longrightarrow R_\star$ \citep{lin2017tidal}. In the limit of vanishing envelope mass, the inertial wave-driven mechanism shuts off. Thus, in our modelling, we suppose that tidal dissipation is negligible when the mass in the convective envelope $M_{CE}\equiv M_\star-M_c$ drops below a critical fraction $\beta_{\rm crit}$ of the total stellar mass. Lacking an {\it a priori} model that would specify this critical fraction, we adopt $\beta_{\rm crit}\sim 3\times 10^{-3}$ based on the modeling and comparison with data performed by \citet{valsecchi2014tidal}.

Putting it all together, the functional form we adopt is
\begin{align}\boxed{\label{eq: Lin}
    k_{2}^1=\frac{k_2}{Q_{10}}=\begin{cases} \epsilon^2\mathcal{F}_0\bigg(\frac{R_c}{R_\star}\bigg)^5 & \mbox{if }M_{CE}>\beta_{\rm crit} M_\star \\
    0 & \mbox{if } M_{CE}\leq\beta_{\rm crit} M_\star
    \end{cases}},
\end{align}
where $\mathcal{F}_0$ is a constant to be determined (see Section~\ref{sec: scale}). As will become clear in Sections~\ref{sec: spin} \&~\ref{sec: scale}, it is convenient to define the dimensionless variables
\begin{align}\label{eq: def}
    \eta\equiv \frac{\Omega_\star}{\Omega_\odot}\,\,\,\,\,\,\,\,\eta'\equiv \frac{m_p\sqrt{G M a}}{\mathcal{K}_\star M_\star R_\star^2 \Omega_\odot}
\end{align}
and the characteristic timescale
\begin{align}\label{eq: timescale}
    \tau&=\Bigg[\frac{3\Omega_\odot}{4\Omega_{\textrm{K}}} \bigg(\frac{m_p}{M_\star}\bigg)\bigg(\frac{R_\star}{a}\bigg)^8\Omega_\odot \alpha^5\mathcal{F}_0\Bigg]^{-1},
\end{align}
so that the evolution of stellar obliquity due to the obliquity tide may be written succinctly as
\begin{align}\label{eq: theta}
    \frac{d\theta}{dt}\bigg|_{10}=-\frac{\sin(\theta)\cos^2(\theta)}{\tau}~\eta^2\bigg[1+\frac{\eta'}{\eta}\cos(\theta)\bigg].
\end{align}
\begin{figure}
\centering
\includegraphics[trim=0cm 0cm 0cm 0cm, clip=true,width=1\columnwidth]{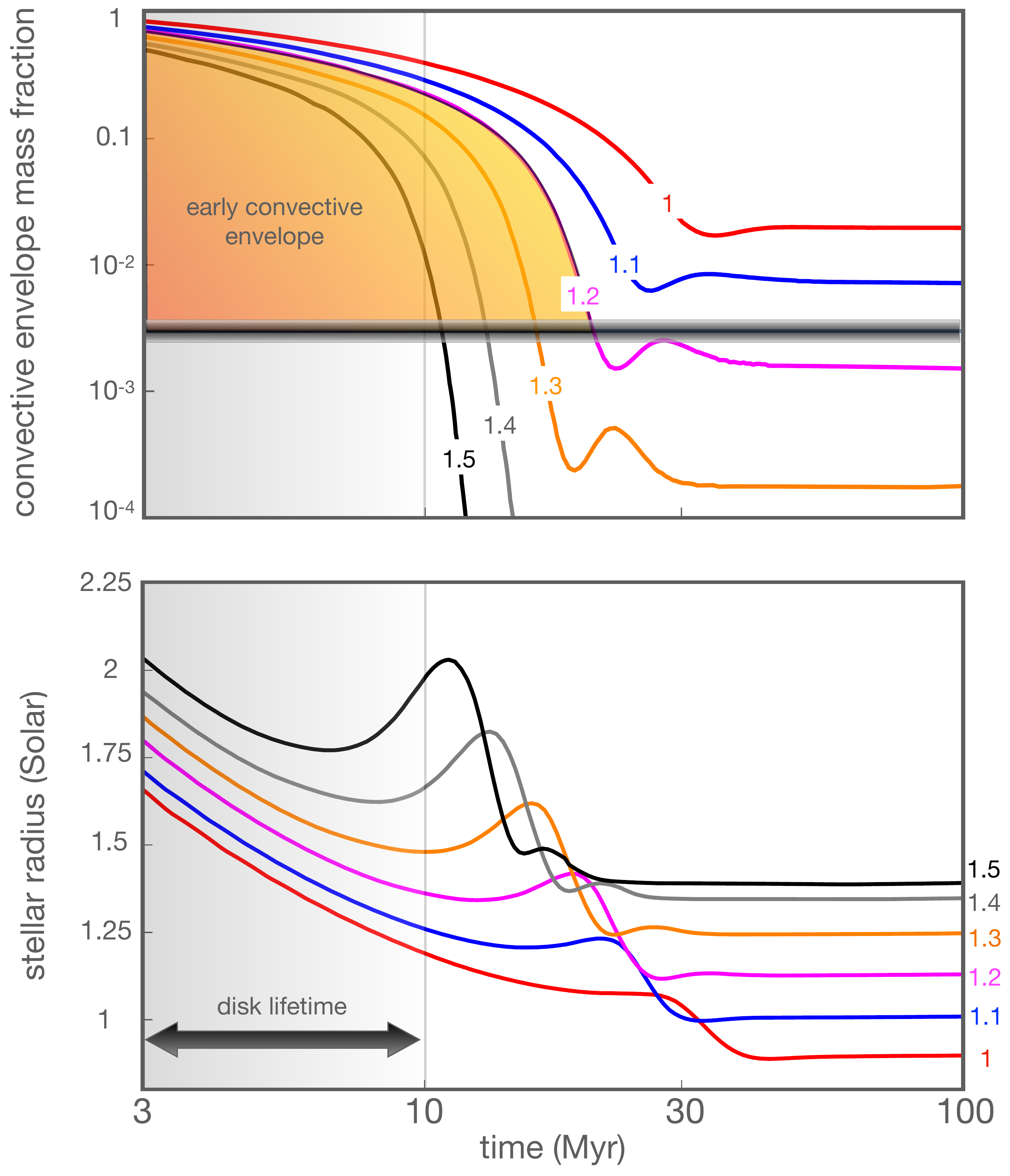}
\caption{Time-evolution of the mass fraction in the convective envelope (top panel) and the stellar radius (bottom panel) for a range of stellar masses. Numbers on lines refer to the stellar mass in units of Solar masses. All stars begin fully convective (top), but those with masses $\gtrsim1.2M_\odot$ lose their convective envelopes after 10s of millions of years (defined here as falling below the horizontal gray line; \citealt{valsecchi2014tidal}). Early-onset stellar obliquities may undergo significant damping during this early, convective phase of more massive stars. Additionally, radii are larger early-on (bottom), further enhancing tidal dissipation.}
\label{fig: early_Envelope}
\end{figure}
\subsection{Equilibrium tide}
In the previous section, we developed the equations for tidal evolution due to the dissipation of inertial waves. The equilibrium tide must also be considered. In terms of an equilibrium value of the tidal quality factor, $Q'_{\textrm{eq}}\equiv Q_{\textrm{eq}}/k_2$, the stellar obliquity and spin rate evolve according to \citep{lai2012tidal}
\begin{align}
   \dot{\theta}\big|_{\textrm{eq}}&=-\frac{3\nu}{2Q'_{\textrm{eq}}}\sin \theta\bigg[\frac{L}{S}-\bigg(\frac{\Omega_\star}{2\Omega}\bigg)\bigg(\frac{L}{S}\cos\theta-1\bigg)\bigg]\\
   \frac{\dot{\Omega}_\star}{\Omega_\star}\bigg|_{\textrm{eq}}&=\frac{3\nu}{2Q'_{\textrm{eq}}}\bigg(\frac{L}{S}\bigg)\bigg[\cos\theta-\bigg(\frac{\Omega_\star}{2\Omega}\bigg)\bigg(1+\cos^2\theta\bigg)\bigg].
\end{align}
In addition, unlike the obliquity tide from inertial wave dissipation, equilibrium tides drive the evolution of semi-major axis according to
\begin{align}
    \frac{\dot{a}}{a}=-\frac{3\nu}{Q'_{\textrm{eq}}}\bigg(1-\frac{\Omega_\star}{\Omega}\cos \theta\bigg).
\end{align}
Efforts to constrain the value and frequency-dependence of $Q_{eq}$ have led to results differing by orders of magnitude \citep{meibom2005robust,penev2012constraining,penev2018empirical}.  For example, recent work on the decay of hot Jupiter orbits has suggested that $Q'_{eq}$ scales to roughly the third power of the tidal forcing frequency \citep{penev2018empirical,anderson2021possible}, taking on values between $10^5$ and $10^7$. While this issue is important, the focus of our work is on the obliquity tide, so we do not present an exhaustive evaluation of the most likely range of values for $Q'_{\textrm{eq}}$. Instead, we adopt $Q'_{\textrm{eq}}=10^6$ as a nominal value unless stated otherwise, and discuss the significance of this choice in section~\ref{sec: polar}.   

The stellar spin axis and planetary orbit respond to the sum of the equilibrium and dynamical tidal evolution, as described above. Following \citet{lai2012tidal}, we write the full evolution equations as: 
\begin{align}\label{eq: full_Eq}
    \frac{\dot{a}}{a}&=\bigg(\frac{\dot{a}}{a}\bigg)_{eq}\nonumber\\
    \frac{\dot{\Omega}_\star}{\Omega_\star}&=\bigg(\frac{\dot{\Omega}_\star}{\Omega_\star}\bigg)_{eq}+\bigg(\frac{\dot{\Omega}_\star}{\Omega_\star}\bigg)_{10}-\bigg(\frac{\dot{\Omega}_\star}{\Omega_\star}\bigg)_{10,\textrm{eq}}\nonumber\\
    \frac{\dot{\theta}}{\theta}&=\bigg(\frac{\dot{\theta}}{\theta}\bigg)_{eq}+\bigg(\frac{\dot{\theta}}{\theta}\bigg)_{10}-\bigg(\frac{\dot{\theta}}{\theta}\bigg)_{10,\textrm{eq}}.
\end{align}
The terms with the subscript $``10,\textrm{eq}"$ take the same functional forms as the obliquity tide equations (\ref{eq: obliquity} \&~\ref{eq: spin-rate}), but with $k_2/Q\big|_{10}$ being replaced by $k_2/Q\big|_{eq}$. The addition of the $``10,\textrm{eq}"$ terms ensures that dissipation due to the $``10"$ component of the tide is not double-counted, given that the equilibrium tide encompasses the responses to all tidal components.

\subsection{Stellar evolution}

Tidal dissipation by way of inertial waves is only possible in the presence of a convective envelope. The thickness of the envelope can be calculated for a given stellar model, which depends on three parameters: mass, metallicity, and age. Solar-metallicity stars with mass $M_\star\gtrsim 1.2 M_\odot$, referred to as the ``Kraft break," have convective zones too thin to dissipate obliquity while on the main sequence. However, even stars above the Kraft break begin their lives fully convective, only losing their convective envelopes after tens of millions of years of evolution. Therefore, if any given hot Jupiter attained its short-period orbit earlier than $\sim$10\,Myr, it would have experienced an early epoch of enhanced inclination-damping around hot and cool stars alike.  Our goal was to determine the conditions in which the early episode of obliquity damping would have damped the obliquities that are observed among modern-day hot Jupiter host stars that lie above the Kraft break. If we find, for instance, that a hot star with a high obliquity should have realigned with its hot Jupiter during the pre-main-sequence phase, we could conclude that (assuming our tidal model is correct) the planet must have attained its short-period orbit after the pre-main-sequence phase.

In order to determine stellar structural parameters as a function of stellar mass, metallicity and age, we downloaded a set of isochrones from the online \textit{SYCLIST} database\footnote{https://www.unige.ch/sciences/astro/evolution/en/database/syclist/}. This database interpolates between the simulation results of \citet{amard2019first} over a stellar mass range between 0.6 and 1.5\,$M_\odot$
and iron metallicities [Fe/H] ranging from $-0.3$ to $+0.3$, and provides output parameters such as $R_\star,\,\mathcal{K}_\star,\,T_{\textrm{eff}}$, in addition to convective zone thickness and mass.\footnote{We use the standard definition [Fe/H]~$=\log_{10}(Z/Z_\odot)$, where $Z$ is the mass ratio of iron to hydrogen.}
 
Figure~\ref{fig: early_Envelope} shows the time evolution of a solar-metallicity star's radius and convective envelope mass fraction, for some representative choices of stellar mass. In the upper panel, the orange shaded region indicates where the mass in the convective envelope exceeds
the critical fraction $3\times 10^{-3}$ below which we have assumed inertial wave damping is deactivated (see Equation~\ref{eq: Lin}; \citealt{valsecchi2014tidal}). 
Young stars also have enlarged stellar radii (by a factor of $1.5-2$) relative to their main-sequence radii (Figure~\ref{fig: early_Envelope}, bottom panel). These expanded radii are highly important in our calculation because of the steep dependence (eighth power) of the
obliquity-damping timescale upon stellar radius. Enlarging the radius of a star by a factor of $1.5$ will cause a star to lose its obliquity $26$ times faster. 

Summarizing these calculations, stars less massive than about 1.2\,$M_\odot$ have convective zones exceeding 0.3\% of their total mass
all the way through to the main sequence, while more massive stars only satisfy this condition for tens of millions of years.
The star's metallicity also affects the thickness of the convective zone, with more metal-rich stars possessing deeper convective envelopes for a given stellar mass \citep{amard2020evidence}. We will discuss this point in Section~\ref{sec: metallicity}. For now, we turn to the evolution of the stellar spin, which also depends upon the thickness of the convective zone, and directly influences the magnitude of tidal dissipation. 

\begin{figure}
\centering
\includegraphics[trim=0cm 0cm 0cm 0cm, clip=true,width=1\columnwidth]{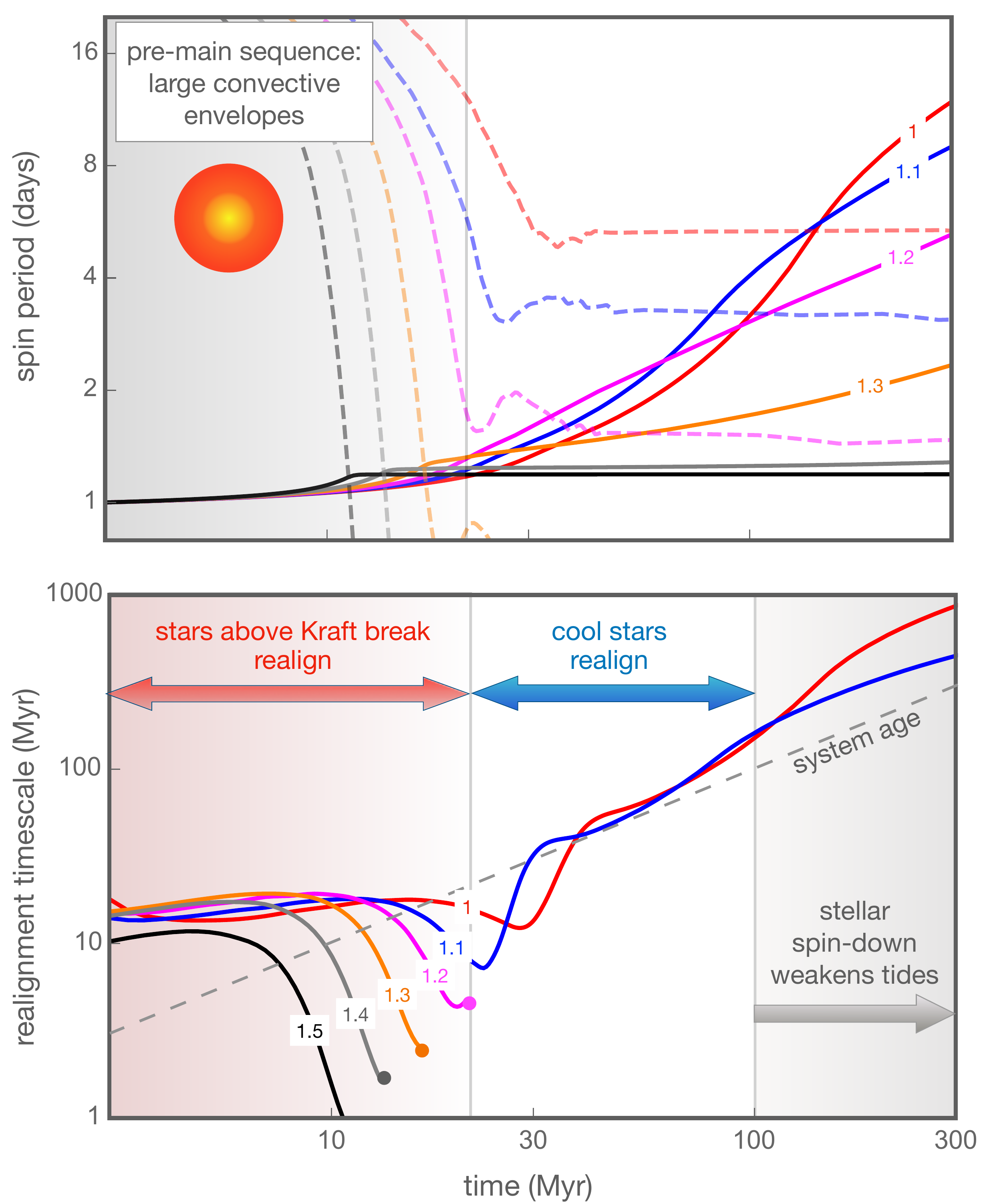}
\caption{Time evolution of the stellar rotation period due to magnetic braking alone (top panel) and the characteristic realignment timescale (bottom panel). In the top, stellar spin periods are denoted by solid lines, and dashed lines track the saturation period, below which magnetic braking becomes less efficient (Equation~\ref{eq: saturation}; \citealt{matt2015mass}). In the bottom panel, the dashed gray line follows the system's age. For stars with $M_\star \geq 1.2 M_\odot$, realignment timescales are short prior to $\sim 10-20\,$Myr, indicative of their thick convective envelopes and large radii. Less massive stars retain short realignment timescales for 100s of millions of years, but eventually spin down enough to drive the realignment timescale above the system's age. Thus, massive stars may realign prior to $\sim20\,$Myr, whereas lower-mass stars may need to realign prior to $\sim200$\,Myr. }
\label{fig: spins}
\end{figure}
\newpage
\section{Stellar spin evolution}\label{sec: spin}

The rate of tidal dissipation due to inertial waves depends quadratically upon stellar spin rate \citep{ogilvie2013tides}. Accordingly, it is important to model stellar spin-down due to magnetic braking concurrently with tidal evolution. Stellar spin-down is thought to be driven by a coupling between the stellar wind and the stellar magnetosphere, causing efficient loss of angular momentum. We adopt the magnetic braking model of \citet{matt2015mass}, in which the crucial parameter is the Rossby number
\begin{align}
    {\rm Ro}\equiv\frac{1}{\Omega_\star \tau_{\rm cz}}.
\end{align}
Here, $\tau_{cz}$ is the convective overturn timescale, one of the outputs of the stellar structure models of \citet{amard2019first}. The value of $\tau_{cz}$ depends mainly on effective temperature \citep{cranmer2011testing,amard2019first}, with secondary dependencies on mass and metallicity. Generally, thinner convective envelopes have shorter convective turnover times. 
For higher stellar rotation rates, the braking efficiency increases until the Rossby number drops below a critical value of order unity, associated with a critical saturated spin rate given by
\begin{align}\label{eq: saturation}
   \Omega_{\textrm{sat}}\approx \chi_{\rm crit} \Omega_\odot \bigg(\frac{\tau_{cz,\odot}}{\tau_{cz}}\bigg),
\end{align}
where $\chi_{\rm crit}\approx10$ following \cite{matt2015mass}. We suppose that stellar spin-down obeys the equation
\begin{align}
    \frac{d\Omega_\star}{dt}=-\frac{T}{\mathcal{K}_\star M_\star R_\star^2}
\end{align}
where the torque $T$ is given by
\begin{align}
    T=T_0\bigg(\frac{R_\star}{R_\odot}\bigg)^{3.1}\bigg(\frac{M_\star}{M_\odot}\bigg)^{0.5}\bigg(\frac{\tau_{cz}}{\tau_{cz,\odot}}\bigg)^p\nonumber\\
    \times\begin{cases} \bigg(\frac{\Omega_\star}{\Omega_\odot}\bigg)^{p+1} & \mbox{if }\Omega<\Omega_{sat} \\
    \bigg(\frac{\Omega_{sat}}{\Omega_\odot}\bigg)^{p}\bigg(\frac{\Omega_\star}{\Omega_\odot}\bigg) & \mbox{if } \Omega<\Omega_{sat}.
    \end{cases},
\end{align}
Choosing $p=2$ reproduces the well-establish \citet{skumanich1972time} law that $\Omega_\star \propto t^{-1/2}$ at late times, and the observed distribution of stellar spin periods across a range of cluster ages is well-matched by setting the torque factor $T_0=9.5\times 10^{30}\,$erg \citep{matt2015mass}. 

With this magnetic braking model, the time evolution of the dimensionless spin rate $\eta$ (defined in Equation~\ref{eq: def}) is given by

\begin{align}\label{eq: eta}
    \frac{d\eta}{dt}=-\gamma \min\big[\eta^2,\eta_{sat}^2\big]\eta\,\,\,\,\,\,\,\, \gamma\equiv \frac{T_0 Y_{\star}}{\mathcal{K}_\star\Omega_\odot},
\end{align}
where we have defined the structural prefactor

\begin{align}
    Y_\star\equiv \bigg(\frac{R_\star}{R_\odot}\bigg)^{3.1}\bigg(\frac{M_\star}{M_\odot}\bigg)^{0.5}\bigg(\frac{\tau_{cz}}{\tau_{cz,\odot}}\bigg)^2,
\end{align}
and $\tau_\odot =12.9\,$days is the convective overturn timescale of the Sun.

For illustration, Figure~\ref{fig: spins} shows how the spin period evolves over time due to magnetic braking alone for solar-metallicity stars of various masses. Early on, the spin periods ($P_\star$, solid lines) are typically below their associated saturation thresholds (dashed lines), leading to exponential spin-down. Then, with time, the saturation period drops as the convective envelope thins. For stars below the Kraft break (blue and red lines), spin-down is significant even after hundreds of Myr.

\section{Magnitude of dissipation}\label{sec: scale}

Next, we determine the order-of-magnitude of $\mathcal{F}_0$, the prefactor setting the magnitude of obliquity-damping due to inertial waves (Equation~\ref{eq: Lin}), by requiring obliquity damping to be efficient for cool stars with hot Jupiters in close orbits, to be consistent with observations. Suppose that the stellar spin rate is faster than the saturation rate, as is typical for cooler stars at early times (Figure~\ref{fig: spins}). The stellar spin will evolve according to (see Equation~\ref{eq: eta})

\begin{align}\label{eq: approx_eta}
    \eta(t)\approx \eta_0 \exp\bigg(-\frac{t}{\tau_{sp}}\bigg)
\end{align}
where $\tau_{sp}\equiv 1/\gamma \eta_{sat}^2$, which we will assume to be constant for illustrative purposes.
We also assume that the stellar obliquity $\theta$ is small and $L/S\ll1$. Substituting this solution into Equation~\ref{eq: theta}, we find

\begin{align}
    \frac{d\theta}{dt}\approx -\frac{\theta}{\tau}\eta_0^2 \exp\bigg(-\frac{t}{\tau_{sp}}\bigg).
\end{align}
By integrating from $t=0$ to $\tau_{sp}$ we obtain the approximate factor by which the obliquity is damped:

\begin{align}\label{eq: approx_Spin}
   \frac{ \theta_{sp}}{\theta_0}\sim\exp\bigg(-\frac{\eta_0^2 \tau_{sp}}{2\tau}\bigg).
\end{align}
To be consistent with observations, we require $\theta_0/\theta_{sp}=10$ when $a_p=8R_\star$ \citep[see, e.g.,][]{dai2017oblique}.
Using our stellar evolution models at age 100\,Myr, and solving equation~\ref{eq: approx_Spin} for a star with an initial spin period
of one day, we find that $\mathcal{F}_0\approx 0.1$.

In comparison, \citet{lin2017tidal} found $\mathcal{F}_0\sim 0.1$ for a homogeneous sphere with a solid core, and $\mathcal{F}_0\sim 0.01$ for a polytrope of polytropic index 1. Thus, the value of 0.1 we require in our model is at the upper end of the numerically-derived range.

\begin{figure}
\centering
\includegraphics[trim=0cm 0cm 0cm 0cm, clip=true,width=1\columnwidth]{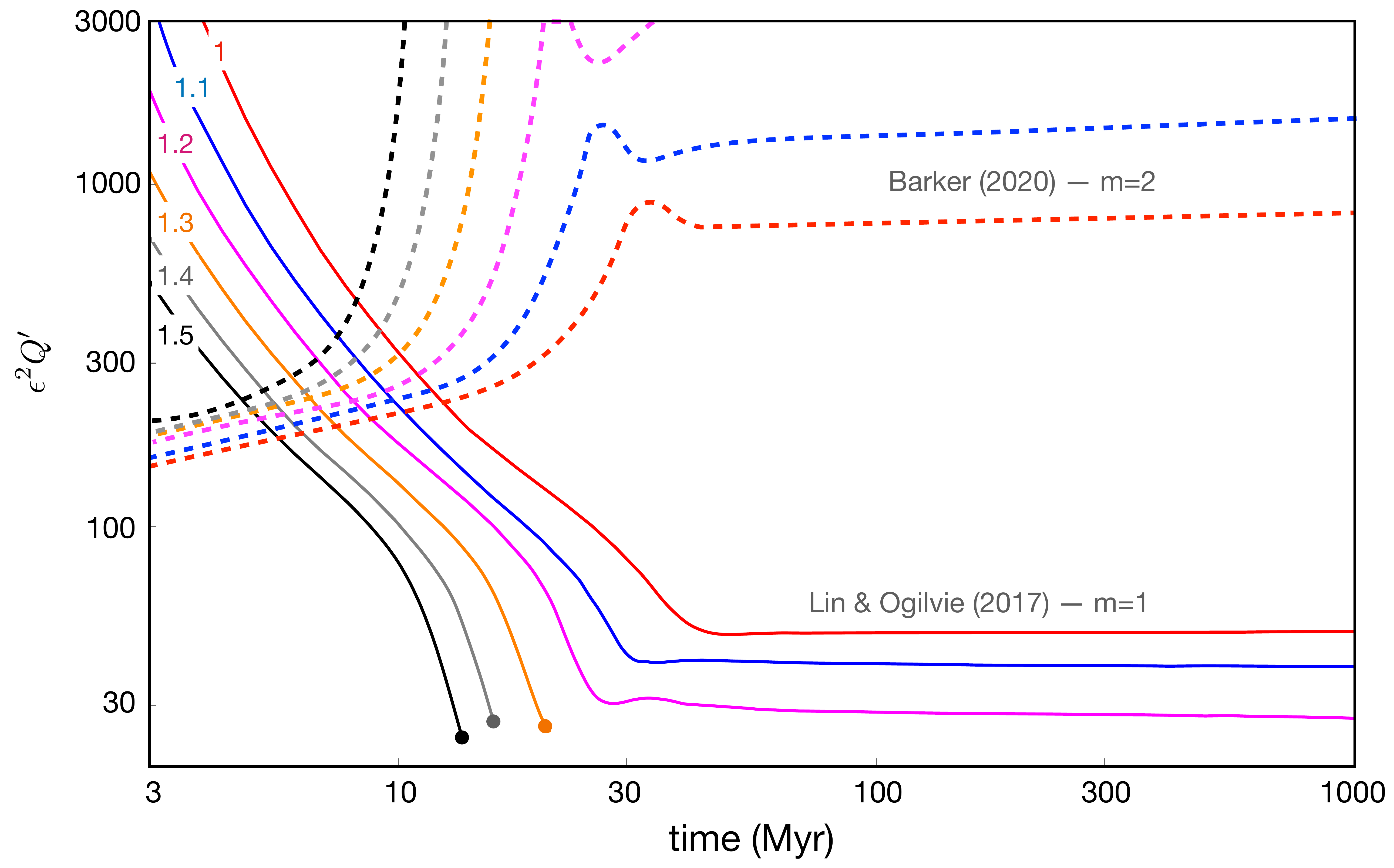}
\caption{Comparison of the tidal dissipation rates calculated by \cite{lin2017tidal} (used in this study) to those computed by \citet{barker2020tidal}, in both cases for solar metallicity. The models predict similar values for the average dissipation rate at 5--10\,Myr, the cruical period for our study,
although the the time derivatives are reversed. After
100\,Myr, the models differ by more than an order of magnitude, but at this late stage, the stellar rotation period is long enough in cool stars to effectively suppress the obliquity tide relative to other tidal mechanisms (see Figure~\ref{fig: spins})}
\label{fig: barker}
\end{figure}

As noted above, \citet{barker2020tidal} carried out numerical simulations of inertial wave damping using realistic stellar structure models, but only for the $m=2$, $l=2$ modes and not the $m=1$ obliquity tide.  Their results disagree with the scaling presented by \citep{lin2017tidal}.
Figure~\ref{fig: barker} illustrates this disagreement by showing the effective magnitudes of dissipation for a range of stellar masses and ages, according to both sets of calculations. The two models exhibit opposite trends during the first 30\,Myr, which corresponds to the the pre-main-sequence phase when the stars gradually lose their convective envelopes (after which stars with $M\gtrsim 1.2M_\odot$ lose them entirely).

The disagreement is probably due to a combination of factors. First, the simulations of \citet{barker2020tidal} considered realistic stellar density profiles, which possessed substantially less mass in their outer convective regions than the homogeneous models of \citet{lin2017tidal}. Intuitively, one might expect that less mass in the convective zone would lead to a lower dissipation rate. However, the energy dissipation rate typically occurs in proportion to the surface area of the radiative core from which the waves themselves are launched \citep{goodman2009dynamical}, which is why the dissipation rate rises with decreasing convective mass, counter to intuition.

Second, \citet{barker2020tidal} and \citet{lin2017tidal} differed in their treatments of the perturbing forces. \citet{barker2020tidal} considered $m=2$ and averaged over a range of perturbing frequencies. \citet{lin2017tidal} did not average over a range of frequencies; they examined the specific frequency of the obliquity tide.
Despite the discrepancies in the time evolution, at least the time-averaged dissipation predicted by the two models agree to within an order of magnitude during the earliest 30\,Myr. Also, in both models, stars with $M_\star\gtrsim 1.2\,M_\odot$ lose their ability to damp obliquities after about 30\,Myr. For less massive stars at late times, the models of \citet{lin2017tidal} predict much higher dissipation than those of \citet{barker2020tidal} --- but, by then, the stars have undergone substantial spin-down which weakens tidal damping.

Although our choice to adopt the scaling of \cite{lin2017tidal} with $\mathcal{F}_0\sim 0.1$ was justified above, it is clear that there is significant room for improvement in understanding the tidal dissipation of inertial waves. 

\section{Tidal erasure simulations}

In this Section, we simulate the obliquity evolution of hot-Jupiter hosts with masses ranging from 1 to 1.5\,$M_\odot$. We first consider systems that are misaligned and prograde. Within this regime, we evaluate two cases:
early arrival of the hot Jupiter (before 3\,Myr), and late arrival (after 100\,Myr). 
Next, we perform analogous simulations for stars that are intially retrograde ($\theta>90^\circ)$. After these illustrative simulations, in Section~\ref{sec: observed} we consider the sample of real stars with measured obliquities to see if early arrival is a viable possibility in our model. Our goal is to identify a subset of hot stars hosting misaligned hot Jupiters that could only have retained their observed misalignments if the hot Jupiters arrived late.

\subsection{Model systems: prograde misaligned}\label{sec: time_Variation} 

We compute the obliquity evolution of stars hosting hot Jupiters at $a_p=0.035\,$AU, for six different stellar masses: $M_\star=\{1,\,1.1,\,1.2,\,1.3,\,1.4\,,1.5\}M_\odot$. We solve Equations~\ref{eq: full_Eq} for $\Omega_\star$, $\theta$ and $a_p$, with an initial spin period of 1\,day at $t=3\,$Myr. We integrate from an initial time $t=t_0$ to a final time of 1\,Gyr.

\begin{figure}
\centering
\includegraphics[trim=0cm 0cm 0cm 0cm, clip=true,width=1\columnwidth]{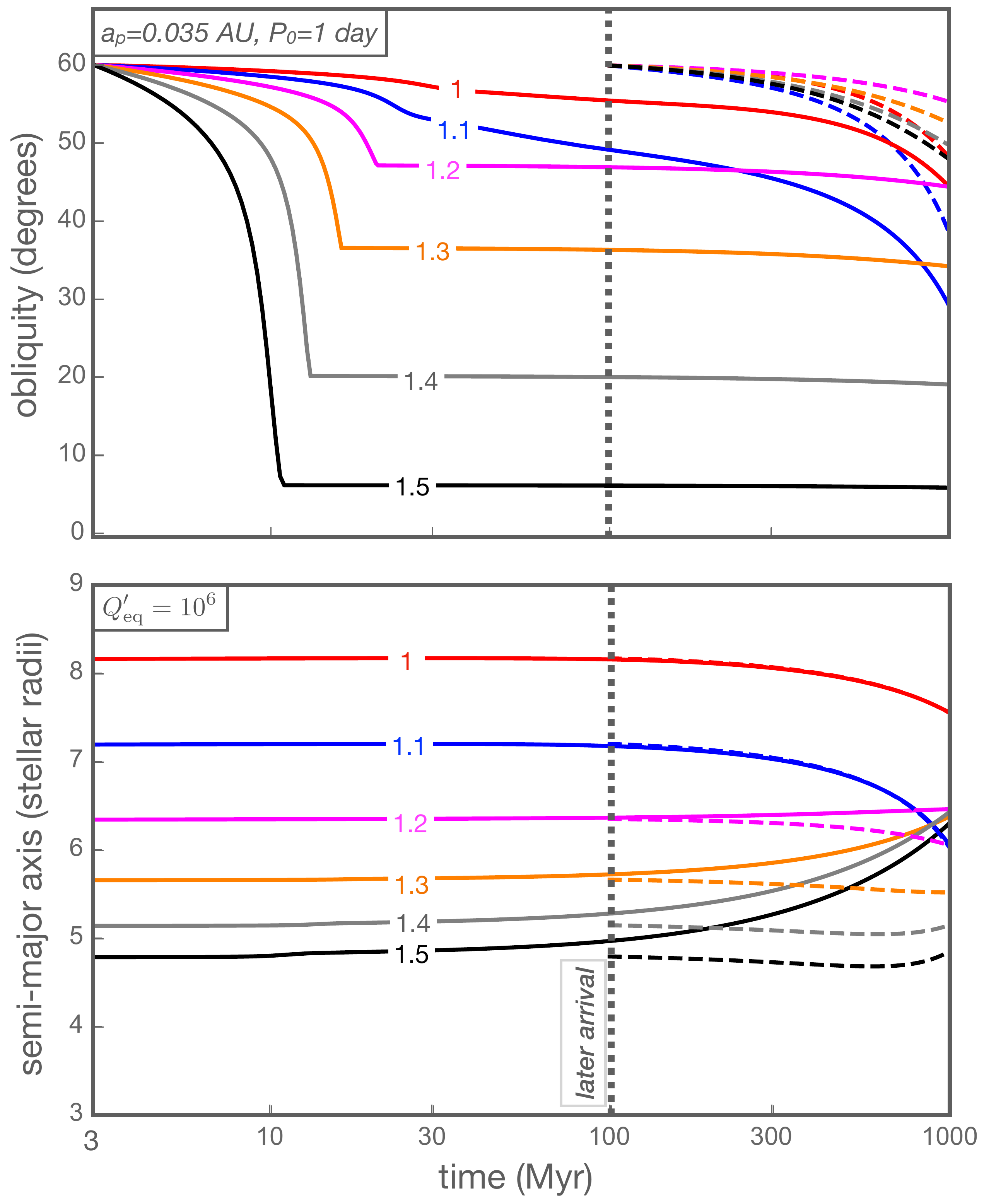}
\caption{Time evolution of the obliquity (top panel, in degrees) and semi-major axis (bottom panel, in units of the stellar radius at 1\,Gyr) of a star with a Jupiter-mass planet at 0.035\,au and an initial obliquity of 60$^\circ$.
Both early arrival (3\,Myr) and late arrival (100\,Myr) are considered. Stars that will be above the Kraft break on the main sequence possess thick enough convective zones at early times to substantially damp the obliquity.}
\label{fig: time1}
\end{figure}

We first consider prograde systems, choosing an initial stellar obliquity of $\theta_0=60^\circ$. Two different choices of $t_0$ are considered: $t_0=3\,$Myr represents early arrival, and $t_0=100\,$Myr represents late arrival. We display the results in Figure~\ref{fig: time1}, the top panel of which shows obliquity evolution and the bottom indicates semi-major axis evolution in units of the star's radius at 1\,Gyr.

Generally speaking, early-arriving hot Jupiters experience a significant amount of realignment over the entire stellar mass range. However, the realignment is more complete within the first 30\,Myr for the higher mass stars. After the pre-main sequence, stars above the Kraft break experience little realignment, while the cooler stars continue to lose their obliquities \citep{li2016tidal}. 

In contrast, obliquity decay for planets arriving late around stars above the Kraft break is driven only by equilibrium tides, due to the lack of convective envelope. Obliquity decay from equilibrium tides is slower than it is from inertial wave dissipation, and it is directly coupled to semi-major axis evolution.  

As noted above, the inclusion of equilibirum tides adds extra degrees of uncertainty to the problem. We chose $Q'_{\textrm{eq}}=10^6$, but the actual value could easily differ by an order of magnitude. Nevertheless, these calculations indicate the potential for initially prograde stars above the Kraft break to undergo substantial realignment \textit{before} reaching the main sequence and losing their convective envelopes. Indeed, prograde rotation causes $a_p$ to initially increase, for early-arriving hot Jupiters around stars more massive than $1.2M_\odot$.

\subsection{Model systems: retrograde \label{sec: retrograde}}

In the previous section, we modelled systems where the stellar obliquity is substantial, but below $90^\circ$. However, a large number of hot stars host hot Jupiters following retrograde orbits \citep{winn2010hot,albrecht2012obliquities}. Indeed, recent work has suggested that a preponderance of oblique hot stars lie close to, but slightly in excess of, $90^\circ$ of obliquity \citep{albrecht2021preponderance}. How do these ``upside-down" systems fit into the picture? 

The obliquity tide's contribution to obliquity decay ($\dot{\theta}_{10}$) vanishes as $\theta\rightarrow 90^\circ$, suggesting that tidal erasure due to inertial wave dissipation is liable to stall at nearly-polar configurations \citep{rogers2013tidal}. Nevertheless, as shown by \citet{li2016tidal}, equilibrium tides continue to damp stellar obliquity even in polar configurations, albeit with an associated decay in semi-major axis. Thus, if one waits for long enough, a polar cool star would eventually drop below $90^\circ$ due to equilibrium tides, at which point the obliquity tide would re-activate and help to drive the system toward alignment. The drawback of this scenario is that if the equilibrium tide is too strong relative to the obliquity tide, then the orbit will decay and the planet will be destroyed before the system aligns (discussed further in Section~\ref{sec: polar}).

\begin{figure}
\centering
\includegraphics[trim=0cm 0cm 0cm 0cm, clip=true,width=1\columnwidth]{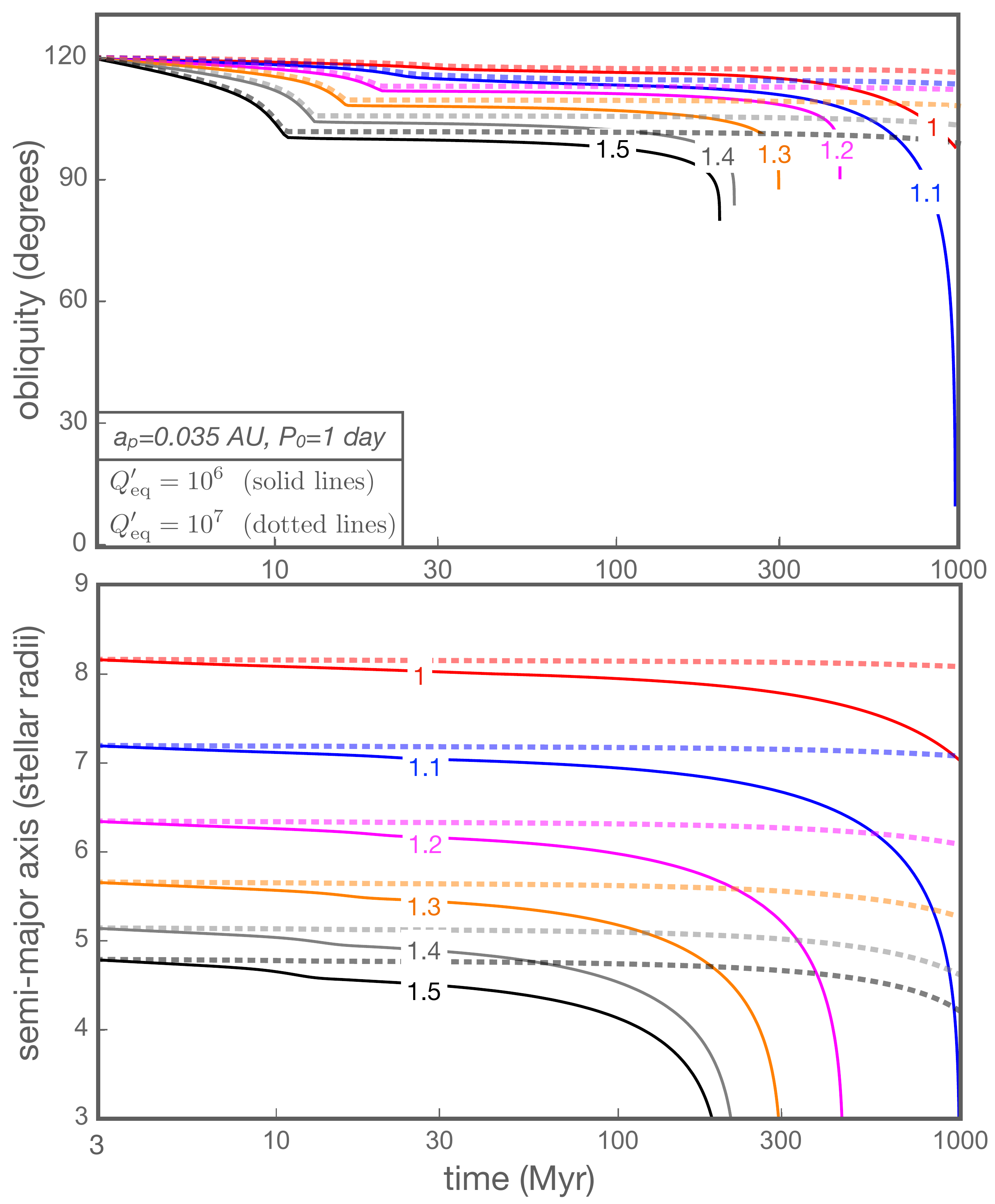}
\caption{Similar to Figure~\ref{fig: time1}, but in this case the initial obliquity is $120^\circ$ and only early-arriving hot Jupiters are modeled. Solid lines correspond to $Q_{\textrm{eq}}=10^6$ while dashed lines are for $Q_{\textrm{eq}}=10^7$. The obliquity tide can  drive higher-mass stars to nearly-polar configurations during the pre-main sequence, but cannot reach prograde configurations due to the slowness of the equilibrium tide. }
\label{fig: time2}
\end{figure}

To illustrate this predicament, we perform similar simulations to the previous section with $\theta_0=120^\circ$ (Figure~\ref{fig: time2}). We only consider early-arriving hot Jupiters, and illustrate two different choices for $Q'_{\textrm{eq}}$: the solid lines are for $Q'_{\textrm{eq}}=10^6$, and the dotted lines are for $Q'_{\textrm{eq}}=10^7$. None of the systems with $Q'_{\textrm{eq}}=10^6$ managed to realign before the planet was engulfed. For the larger choice of $Q'_{\textrm{eq}}$, the obliquities of the more massive stars were lowered
toward 90$^\circ$ before losing their convective envelopes and stalling. 

\subsection{Observed systems}\label{sec: observed}

The preceding calculations, involving fabricated systems, show that obliquity erasure during the pre-main sequence depends sensitively on the orbital distance and the stellar properties. In this section, we consider the observed systems. Our goal is to identify any misaligned hot stellar hosts that \textit{would} have realigned with their planet if the hot Jupiter were present during the pre-main sequence.

We downloaded the stellar and orbital properties of all the systems with measured stellar obliquities from the online TEPCAT database \citep{southworth2011homogeneous}. We identified all the stars lying within the range covered by our stellar structure models ($0.6<M_\star/M_\odot<1.5$ \& $-0.3<\textrm{[Fe/H]}<0.3$). We did not limit the sample in orbital distance or planet mass. These selection criteria lead to a sample of 104 systems.
 
For each system, we extracted the observed planetary properties $m_p$ and $a_p$, together with the stellar quantities $M_\star$ and $Z_\star$ in order to interpolate the time series of all stellar structure parameters that are relevant to our tidal and spin-evolution models (see Eqs~\ref{eq: theta} \&~\ref{eq: eta}). Using these physical parameters, we repeated the obliquity evolution calculation in the previous section, and set the initial stellar obliquity equal to the observed projected stellar obliquity $\lambda$. As in section~\ref{sec: time_Variation}, we performed separate calculations for early arrival (3\,Myr) and late arrival (100\,Myr). We simulated each system until $t=1\,$Gyr and extracted the final obliquity $\theta_f$.

To isolate the effect of an early convective envelope, we ignored the influence of the equilibrium tide in this section ($Q'_{\textrm{eq}}\rightarrow \infty)$. As illustrated in Figures~\ref{fig: time1} \&~\ref{fig: time2}, semi-major axis evolution can be substantial in the first gigayear, but most of the orbital decay occurs close to the end of the planet's life. In a statistical sense, it is unlikely to catch a planet during the final stage of orbital decay, and thus it should be rare to observe a system for which the orbit has decayed substantially.

Nevertheless, the orbit of WASP-12b has been observed to be decaying \citep{maciejewski2016departure,patra2017apparently, yee2019orbit}, possibly triggered by the star's evolution away from the main sequence \citep{weinberg2017tidal} or a high planetary obliquity \citep{millholland2018obliquity}. We leave the analysis of the relative importance of each tidal component for future work and consider only realignment due to obliquity tides. (See \citealt{anderson2021possible} for an approach that incorporates both tidal components into a common formalism.)

\begin{figure*}
\centering
\includegraphics[trim=0cm 0cm 0cm 0cm, clip=true,width=0.9\textwidth]{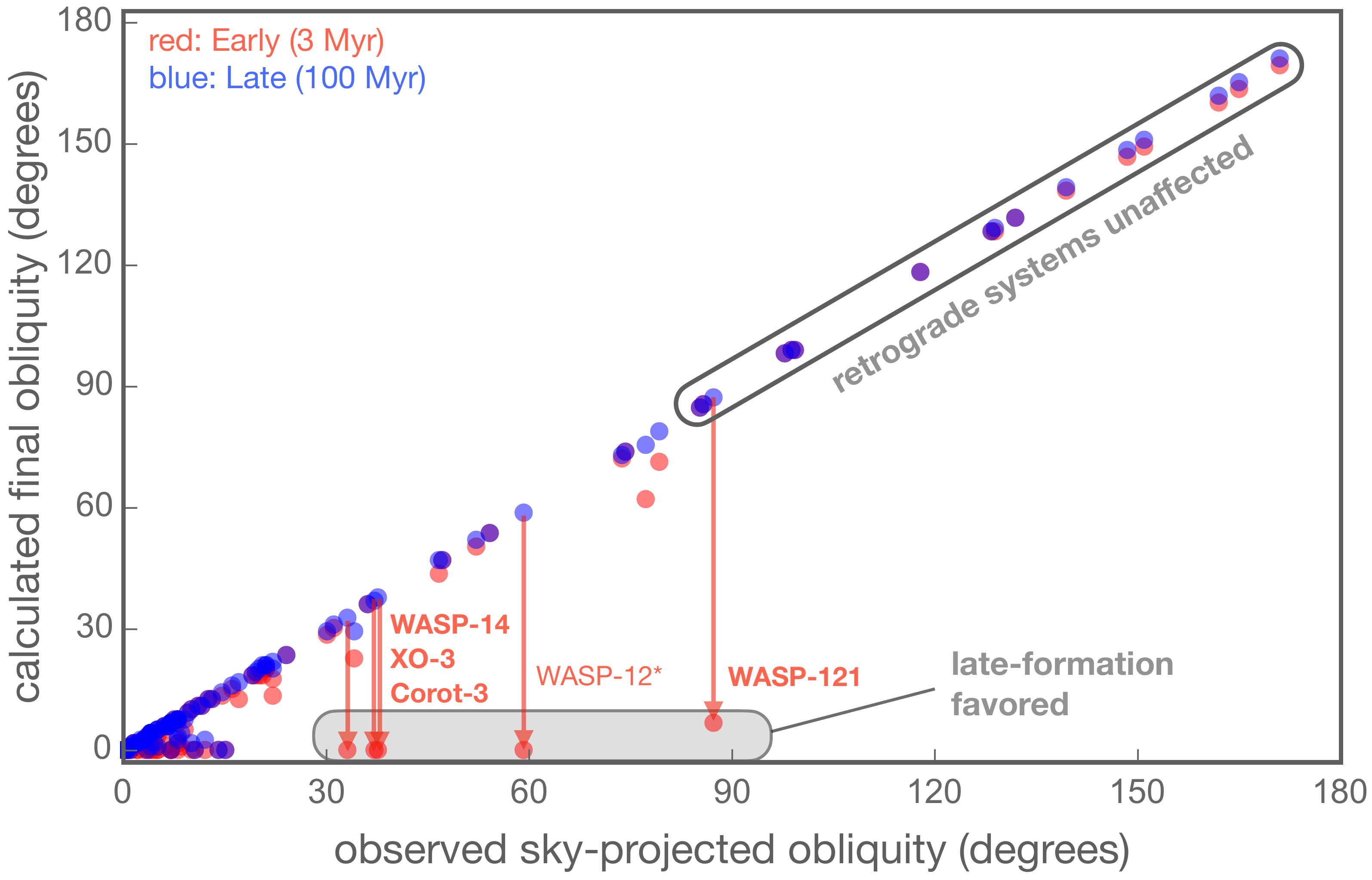}
\caption{The {\it calculated} final obliquity for stars with initial obliquities equal to their \textit{observed} modern-day projected obliquity $\lambda$, plotted as a function of $\lambda$ \citep{southworth2011homogeneous}.Red points represent the early-arrival calculations (3\,Myr) and blue points represent the late-arrival calculations (100\,Myr). In the late-arriving calculations, there are no cases in which the calculated obliquity is lower than $30^{\circ}$ and the observed obliquity is higher than $30^\circ$. In this sense, the late-arrival calculations are consistent with observations. In the early-arriving calculations, there are five systems (with names printed on the figure) for which the calculated obliquity is low and the observed obliquity is high, a contradiction that suggests the planets in these systems arrived late. These cases are discussed in Section~\ref{sec: Late}.}
\label{fig: data_Model}
\end{figure*}

In Figure~\ref{fig: data_Model}, the final calculated obliquities are plotted in the vertical dimension, with red points denoting early-arriving planets and blue points denoting late arrivals. The horizontal dimension shows the corresponding measurement of the sky-projected obliquity. Note first that there are no points in the bottom-right quadrant: none of the systems known to be retrograde ($>90^\circ$) are predicted, in the framework of our tidal model, to have aligned within a billion years. This is because of the weakness of the obliquity tide in polar configurations. Indeed, none of the retrograde systems showed any significant movement towards a polar configuration, indicating that all retrograde systems are tidally decoupled over gigayear timescales. This implies that the observed overabundance of nearly-polar systems \citep{albrecht2021preponderance} cannot be attributed to tidal evolution
according to our model. Rather, this overabundance (if it is confirmed
with future observations) may have been caused by the
obliquity excitation mechanism.

Crucially, there are five systems for which the observed sky-projected obliquity exceeds 30$^\circ$ and yet our model predicts that they should have lost these obliquities if the planet arrived early. These five systems are XO-3b, Corot-3b, WASP-12b, WASP-14b, and  WASP-121b. As mentioned above, WASP-12b's orbit is decaying, implying that its semi-major axis was larger in the past. The other 4 planets must have arrived late in order to have retained a high stellar obliquity. We discuss these systems at greater length in Section~\ref{sec: Late}.

\section{Discussion}

Stars with $T_{\rm eff}\lesssim 6100$K hosting hot Jupiters within about 8 stellar radii tend to have low obliquities, while hotter stars with similar planets have a wide range of obliquities \citep{winn2010hot,albrecht2012obliquities}.
It has been proposed that tidal dissipation within the convective envelopes of cooler stars removes their obliquities
\citep{winn2010hot}. A promising theory for the tidal mechanism relies upon the excitation and dissipation of inertial waves within the convective envelope of the star \citep{lai2012tidal,ogilvie2013tides,lin2017tidal}. 
As we have emphasized, even the stars that are currently hotter than 6100\,K were once fully convective.
The sharp distinction known as the Kraft break only materializes when these stars lose their convective envelopes, which takes tens of millions of years. We have investigated whether this early stage of pre-main-sequence tidal dissipation should have damped the stellar obliquities of currently-hot stars, if the planet's orbit were sufficiently close at an early enough stage.

\subsection{Evidence for late formation}\label{sec: Late}

By using tidal theory in conjunction with stellar evolution models, we identified four high-obliquity hot Jupiters that, within the framework of our model, could not have arrived early. They are XO-3b, Corot-3b, WASP-14b and WASP-121b. Of these, WASP-121b is the most striking. The planet has a nearly-polar orbit \citep{bourrier2020hot}, whereas in our calculation the star's obliquity should have been damped to no more than a few degrees if the planet had arrived early.
WASP-14b is a similar case, although the measured stellar obliquity of
$33.1 \pm 7.4^\circ$ \citep{johnson2009third} is not as high as for WASP-121. Either the hot Jupiters in these systems arrived after the dispersal of the protoplanetary disk,
or the obliquities were excited after arrival (discussed below). We encourage follow-up observations that are capable of detecting wide-orbiting companions that could have initiated high-eccentricity migration in these systems.

As noted previously, WASP-12b is a special case. The calculated final obliquity is much lower than the observed obliquity ($\lambda=59^{+15}_{-20}$$^\circ$; \citealt{albrecht2012obliquities}). This system is unusual in that the interval between transits is shrinking \citep{maciejewski2016departure,patra2017apparently}, likely due to tides in the star \citep{bailey2019understanding,yee2019orbit} or possibly the planet \citep{millholland2018obliquity}. Moreover, the host star's observable properties are compatible with an incipient subgiant star, and the recent structural change of the star may have triggered the rapid orbital decay \citep{weinberg2017tidal,bailey2019understanding}. These factors suggest that the system has undergone significant orbital decay that is not accounted for in our model. Thus, the obliquity decay computed in our model might be an overestimate, and we cannot conclude that WASP-12b formed late.

Two other planets for which the calculated spin-orbit misalignment in the early-arrival scenario is significantly lower than the observed value are Corot-3b ($\lambda=37.6^{+22.3}_{-10}$$^\circ$) and XO-3b ($\lambda=37.3\pm3$$^\circ$; \citealt{addison2018stellar,triaud2009rossiter,hirano2011further}). XO-3b is especially interesting because its larger orbital eccentricity of $\approx$0.3 \citep{bonomo2017gaps} independently suggests a late arrival.

Throughout this work we have implicitly assumed that the stellar obliquity is excited at the same time as the hot Jupiter's arrival onto a short-period orbit. This assumption is appropriate for many of the proposed origin scenarios for stellar obliquity, including the misalignment between the star and its protoplanetary disk \citep{batygin2012primordial,spalding2014early,fielding2015turbulent}, along with high-eccentricity migration \citep{anderson2016formation}. However, it is also possible that a hot Jupiter arrives on a well-aligned orbit, via disk-driven migration (or even through coplanar high-eccentricity migration; \citealt{petrovich2015hot}), and the misalignment occurs later.
The proposed scenarios that have this property all depend upon the internal dynamics of the star, including internal angular momentum transport within the star due to internal gravity waves \citep{rogers2012internal,rogers2013internal}, the fluid elliptical instability \citep{cebron2013elliptical}, and asymmetric angular momentum loss from stellar winds \citep{spalding2019stellar}. Each of these mechanisms depends upon an array of assumptions, the validity of which remains difficult to ascertain. 

Another assumption that we made is that protoplanetary disks do not last longer than 10\,Myr. While this is a good assumption at the population level \citep{haisch2001disk,mamajek2009initial}, some disks are observed to last longer than 10\,Myr \citep{silverberg2020peter}. These longer-lived disks are usually associated with stars that will become M-dwarfs, which rarely host hot Jupiters \citep{dawson2018origins}. Nevertheless, examples exist of more massive stars with longer-lived disks (such as TW Hydrae; \citealt{weinberger2012distance,powell2019new}) such that at least some hot Jupiters may emerge from their disks after their host star's pre-main sequence phase.

\subsection{Critical distance to form early}

Using the arguments outlined above, it is possible to obtain a simple analytic estimate of the critical orbital distance for a
hot Jupiter within which the star would have re-aligned during the pre-main sequence phase. For this calculation, we do not specify a particular tidal dissipation mechanism, but rather work generally in terms of $k_2/Q_{10}$. We begin with equation~\ref{eq: obliquity}, but assume $L/S\ll 1$, giving
\begin{align}
    \frac{d\theta}{dt}&=-\frac{3}{4}\frac{k_2}{Q_{10}}\bigg(\frac{m_p}{M_\star}\bigg)\bigg(\frac{R_\star}{a}\bigg)^5\Omega_{\textrm{K}}\sin(\theta)\cos^2(\theta).
\end{align}
We consider a star with initial obliquity of $\theta_0$ at $t_0=3\,$Myr and integrate to a final time $t=\tau_{\textrm{CE}}$,
the time at which the star loses its convective envelope.
We denote by $\theta_f$ the final obliquity, leading to the expression
\begin{align}\label{eq: limit}
    \int_{\theta_0}^{\theta_f}\frac{d\theta}{\sin(\theta)\cos^2(\theta)}&
    =-\frac{3}{4}\frac{k_2}{Q_{10}}\bigg(\frac{m_p}{M_\star}\bigg)\bigg(\frac{R_\odot}{a}\bigg)^5\Omega_{\textrm{K}} \mathcal{I}(M_\star),
\end{align}
where we have defined the integral
\begin{align}
    \mathcal{I}(M_\star)\equiv\int_{t_0}^{\tau_{\textrm{CE}}(M_\star)}\bigg(\frac{R_\star(M_\star,t)}{R_\odot}\bigg)^5dt,
\end{align}
with explicit dependence upon the stellar mass included. We assume [Fe/H]~$=0$ for simplicity in this calculation.

We numerically interpolated the stellar-evolutionary models to obtain an approximate expression for $\mathcal{I}$:
\begin{align}
    \frac{\mathcal{I}(M_\star)}{\textrm{Myr}}\approx 230 \bigg(\frac{M_\star}{M_\odot}\bigg)-170.
\end{align}
The equations above can be solved for any given values of $k_2/Q_{10}$, $\theta_0$, and $\theta_f$.

Suppose $\theta_0=60^\circ$, and we require $\theta_f=1^\circ$ for the system to be deemed ``aligned". For this case,
the left-hand side of Equation~\ref{eq: limit} has a value of $-5.2$.
If the star is hotter than the Kraft break, we would expect the system to have been misaligned
if the hot Jupiter arrived early at an orbital distance smaller than
\begin{align}\label{eq: criterion}
    \boxed{a_{\rm align}\approx 0.02\Bigg[\bigg(\frac{10^5}{Q'_{10}}\bigg)\bigg(\frac{(M_\star/M_\odot)-0.74}{\sqrt{M_\star/M_\odot}}\bigg)\Bigg]^{2/13}\textrm{AU}},
\end{align}
where we have assumed a Jupiter-mass planet.

The expression above is independent of the inertial-wave dissipation model that we used throughout this paper,
although the nominal value $Q'_{10}=10^{5}$ was taken from this model.
Based on the preceding arguments, any hot Jupiter orbiting a strongly misaligned hot star cannot have
formed via disk-driven migration or \textit{in situ} formation if it orbits closer
than $a_{\rm align}$. Otherwise, the host star's early convective envelope would have
realigned to stellar equator with the orbit.

\subsection{Realignment of polar systems}\label{sec: polar}

In Section~\ref{sec: retrograde} we found that although initially prograde stars can realign during the pre-main sequence, initially retrograde stars generally cannot. Equilibrium tides are required to pass through the polar configuration, and equilibrium tides are thought
to require longer timescales than the pre-main sequence lifetime.

Even over longer timescales, a retrograde system can only reach a well-aligned state if the obliquity tide remains strong. The rate of obliquity damping due to inertial waves scales quadratically with the stellar spin rate \citep{ogilvie2013tides}.
By inspection of Figure~\ref{fig: spins}, stars below the Kraft break have realignment timescales that exceed the system's age
after roughly 100\,Myr of evolution. This suggests that hot Jupiters arriving later than 100\,Myr would not have had enough time to realign with their host stars.

To be more quantitative on this matter, we estimate the required relative strengths of the equilibrium and dynamical tides in order for retrograde systems to realign, either on the pre-main sequence or beyond. Consider equations~\ref{eq: full_Eq} with $\theta\approx \pi/2$, and
define the small variable $\Delta \theta\equiv \theta-\pi/2$. Given that the obliquity tide vanishes at $\Delta \theta =0$, there exists some finite range $|\Delta \theta|\leq\Delta \theta_c$ within which the obliquity decay due to the equilibrium tide exceeds that due to the obliquity tide, i.e., $\dot{\theta}_{\textrm{eq}}>\dot{\theta}_{10}$. By expanding equations~\ref{eq: full_Eq} about $\Delta \theta \approx 0$, we find
\begin{align}
    \dot{\theta}_e&=\dot{\theta}_{10}\nonumber\\
    \frac{1}{2}\frac{1}{Q'_{10}}\Delta \theta^2 &\approx \frac{1}{Q'_{\textrm{eq}}}\bigg[\frac{L}{S}+\frac{\Omega_\star}{2\Omega}\bigg]\nonumber\\
    \rightarrow \Delta \theta_c &\approx \sqrt{2B\, \frac{Q'_{10}}{Q'_{\textrm{eq}}}},
\end{align}
where $B\equiv \frac{L}{S}+\frac{\Omega_\star}{2\Omega} \sim 1$.

We may then suppose that any retrograde star must be pushed through
the obliquity range $|\Delta \theta| < \Delta \theta_c$ by equilibrium tides if it is to reach the prograde-aligned state. Equilibrium tides also cause the semi-major axis to evolve. If the planet is to survive,
the semi-major axis must not undergo substantial decay during the traversal of a polar configuration. To quantify this criterion, we write the evolution of $a$ relative to $\theta$ due to equilibrium tides alone as
\begin{align}
    \frac{da_p}{d\theta}\bigg|_{\textrm{eq}}\approx \frac{2a}{B},
\end{align}
where we have again taken the limit $\Delta \theta\rightarrow 0$. Solving this equation for the region $|\Delta \theta| < \Delta \theta_c$, we find that the semi-major axis decreases to a fraction
\begin{align}
\label{eq:orbitaldecay-factor}
    \frac{a_p}{a_0}\approx \exp\bigg(-4\sqrt{\frac{2\,Q'_{10}}{B\,Q_{\textrm{eq}}}}\bigg)
\end{align}
of its initial value.

To gauge the importance of the preceding constraint, suppose we require that $a_p$ changes by less than 10\% ($a_p>0.9 a_0$). Solving equation~\ref{eq:orbitaldecay-factor} with $B\sim 1$ leads to the
requirement $Q'_{10}\lesssim 10^{-3.5}\,Q'_{\textrm{eq}}$. Loosely speaking, the obliquity tide must exceed the equilibrium tide by a factor of at least $10^3$ or else retrograde-orbiting planets will be destroyed before becoming well-aligned (this condition was satisfied in the calculations by \citealt{li2016tidal}). 

By inspection of Figure~\ref{fig: barker}, tidal dissipation from inertial waves yields $Q'_{10}\sim 10^{2-3}/\epsilon^2\approx 10^{4-5}(P_\star/\textrm{day})^2$  \citep{lin2017tidal,barker2020tidal}. Thus, in order for retrograde systems to realign according to the inertial wave theory, cool stars would need to retain rapid rotation long into their main sequence lives \textit{and} the equilibrium tide quality factor
must exceed $Q'_{eq}\gtrsim 10^{7-8}$.

There is observational evidence that tides cause the host stars of hot Jupiters to spin faster \citep[see, e.g.,][]{tejada2021further}. Indeed, once the stellar rotation period lengthens beyond tens of days, tidal spin-up can overcome magnetic braking, before entirely quenching stellar spin-down \citep{penev2018empirical}. This effect erases the system's memory of the original stellar spin rate, making the star appear younger. Such a late phase of rapid rotation may potentially extend the interval of time during which obliquity tides are dominant. 

For the parameters chosen in our work, the equilibrium tide is generally too strong relative to the obliquity tide to accomplish the realignment of retrograde systems. Realignment could feasibly occur if the strength of the obliquity tide were drastically raised \citep{damiani2018influence}, but our choice of $\mathcal{F}_0=0.1$ is already at the upper end of numerical estimates \citep{lin2017tidal,barker2020tidal}. On the other hand, if the magnitude of dissipation from inertial waves is higher than assumed here, we could identify a larger number of hot Jupiters that must have arrived late. For example, remaking Figure~\ref{fig: data_Model} using $\mathcal{F}_0=1$ yields twice as many systems with observed misalignments that would have realigned during the pre-main sequence.  

\begin{figure*}
\centering
\includegraphics[trim=0cm 0cm 0cm 0cm, clip=true,width=0.8\textwidth]{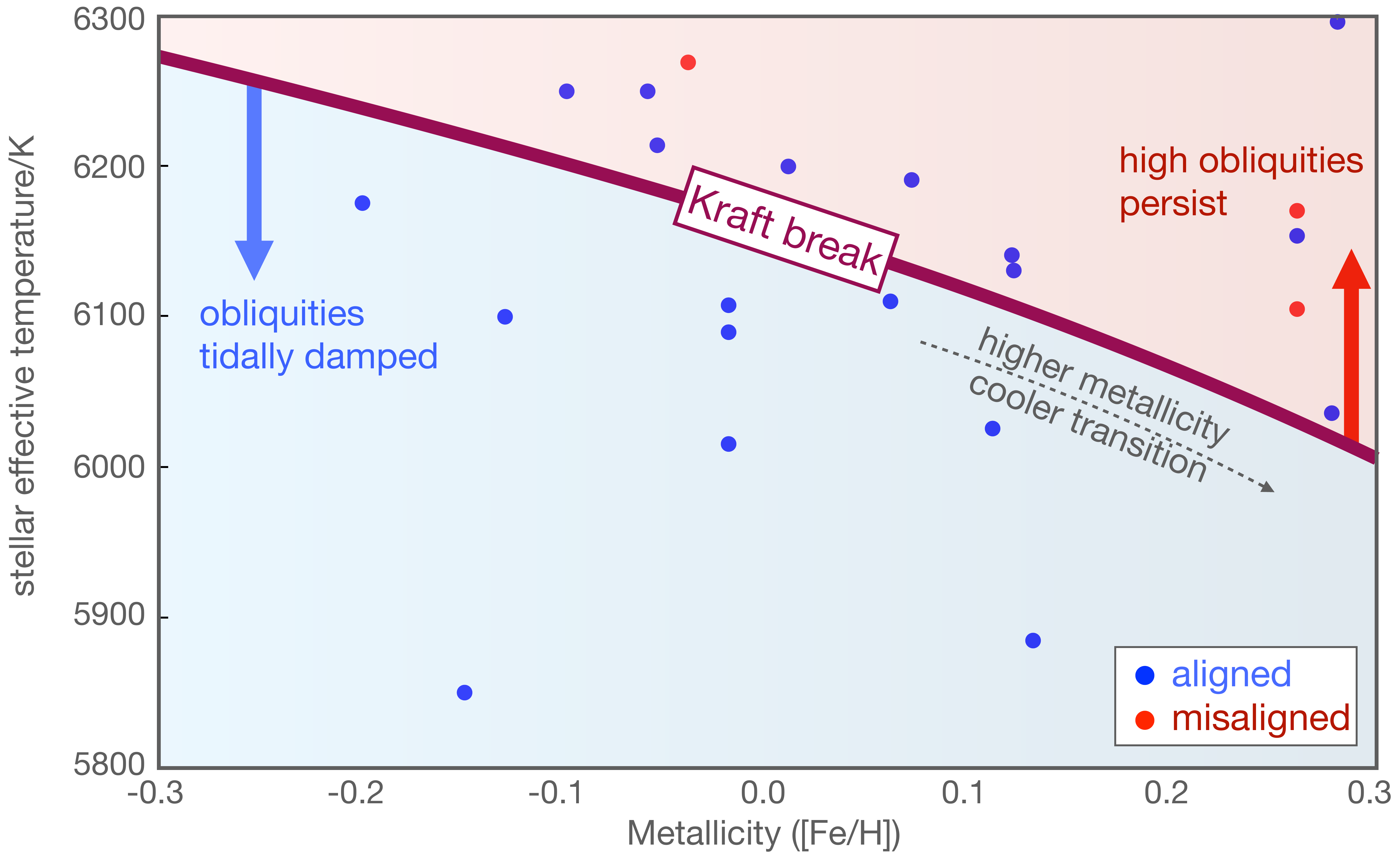}
\caption{Effective temperature demarcating the Kraft break, for various metallicities. Higher metallicity leads to deeper convective envelopes. Consequently, when stars lose their convective envelopes, the convective mass of a high-metallicity star is higher than that of a low-metallicity star. Higher metallicity stars are also cooler for a given stellar mass. Together these effects cause the Kraft-break temperature to decrease with increasing metallicity.}
\label{fig: metals}
\end{figure*}

Cumulatively, it appears difficult to explain the well-aligned configurations of cool stars using the theory of inertial waves, unless they were intially prograde, or inertial wave dissipation is stronger than assumed here. This discrepancy is separate from our primary focus here, i.e., whether massive stars can realign during the pre-main sequence. But nonetheless it is a related issue and should be addressed in future work.

\subsection{Dependence upon metallicity}\label{sec: metallicity}

Our simulations hinge upon the assumption that tides sculpt the obliquity trends of hot Jupiter hosts. We have tried to think of testable model predictions arising from the tidal erasure hypothesis. The most promising test is a trend with metallicity. Specifically, the transition between aligned cool stars and misaligned hot stars is a sharp function of effective temperature (see Figure~\ref{fig: Data1}). If the tidal model utilized here is to be believed, the transition must correspond to the point at which stars lose their convective envelopes.

This transition is not only a function of mass, but also metallicity and age (see Figure~\ref{fig: early_Envelope}). This point was highlighted by \cite{amard2020impact}, who found evidence for a metallicity dependence of the spin rates of stars observed by the Kepler mission \citep{amard2020evidence}. They found that when all other observed characteristics are roughly the same, higher-metallicity stars spin faster. We propose that an analogous trend should be found in the stellar obliquity data. The critical value of $T_{\rm eff}$ above which the obliquity distribution broadens substantially should \textit{decrease} with increasing metallicity (Figure~\ref{fig: metals}).

According to the stellar-evolutionary models, when two stars have equal masses, the more metal-rich star
has a higher global opacity \citep{amard2020impact}, leading to a lower effective temperature,
and a deeper convective envelope. To illustrate this point, we consulted the models for varying metallicity and found the $T_{\rm eff}$ for which the convective mass is $\beta\sim3\times 10^{-3}$ of the total stellar mass \citep{valsecchi2014tidal}. Figure~\ref{fig: metals} displays this ``Kraft temperature" as a function of metallicity. This is the basis of the prediction that the Kraft temperature decreases with increasing metallicity.

Superimposed on Figure~\ref{fig: metals} are the known planets with $M_p>0.3\,M_J$ and $a_p<8\,R_\star$ for which the stellar obliquity has been measured. Of these systems, those that are misaligned by more than $\lambda=30^\circ$ with a statistical confidence greater than 2-$\sigma$ are shown in red \citep{southworth2011homogeneous}. The rest are in blue. Our prediction requires that no red points should exist below the curve in Figure~\ref{fig: metals}. This is true within the current sample.

We look forward to ongoing measurements refining this prediction. More data are needed for stars with a narrow range of masses, a wide range of metallicities, and effective temperatures spanning the range from 6000 to 6300K. This will not be easy, not only because of the narrow restrictions, and the need for accurate and precise values of the effective temperature, but also because hot Jupiters are rare around low-metallicity stars.  

\section{Conclusions}

Stars hotter than the ``Kraft break" ($T_{\rm eff}\gtrsim 6100$\,K) hosting hot Jupiters often have high obliquities (Figure~\ref{fig: data_Model}). This trend is thought to arise from strong tidal dissipation in the convective envelopes of cool stars that erase their obliquities. In this work, we have extended the tidal picture to include the pre-main sequence evolution of stars above the Kraft break. Specifically, these stars possess thick convective envelopes during their earliest $\sim$20\,Myr of evolution before losing them (Figure~\ref{fig: early_Envelope}). Thus, any early-arriving hot Jupiters around such stars would drive an early phase of tidal obliquity removal analogous to that experienced by cooler stars on the main sequence \citep{ogilvie2013tides, amard2019first}. 

We derived a minimum separation that a hot Jupiter must possess in order to avoid pre-main sequence realignment (Equation~\ref{eq: criterion}), which lies at roughly $0.02\,$AU, though it depends upon the stellar mass and tidal parameters chosen. We identified a set of 4 highly-inclined hot Jupiters orbiting hot stars that, within the context of our model, would have lost their spin-orbit misalignments had they formed early. These systems include XO-3b, Corot-3b, WASP-14b \& WASP-121b, and they likely obtained their short-period, inclined orbits subsequent to a few $10s$ of millions of years. Future refinements to the tidal theory will likely alter this list of special systems. The perturbations from as-yet unseen external companions, or long-lived natal disks, are possible drivers of late-arriving hot Jupiters.

Systems that were born retrograde are generally unable to realign during the pre-main sequence phase, whether the star is above or below the Kraft break. Obliquity tides cause the evolution to stall when the star has a nearly-polar orientation. The slower action of equilibrium tides is required to push the star into a prograde state. Qualitatively, this
sounds like a possible explanation for the observed overabundance of polar systems \citep{albrecht2021preponderance}, but in our simulations the currently-retrograde systems are tidally decoupled. This suggests that their polar configurations are a result of formation, rather than tidal evolution.

Additionally, cool stars spin down over time, which weakens the tidal damping of obliquity \citep{ogilvie2009tidal,ogilvie2013tides}. Eventually, the timescale over which stellar obliquity is damped rises above the system's age (Figure~\ref{fig: spins}). Thus, if hot Jupiters arrive around cool stars with large inclinations later than this epoch, they either decay into the star before realigning, or retain their misalignments, conflicting with the observed alignment of cool stars. The combined uncertainties of stellar spin down and tidal damping prevent us from placing strong constraints upon this critical age, but we suggest tidal realignment becomes ineffective after several 100s of millions of years. 

Finally, the tidal hypothesis requires that the transition between aligned cool stars and misaligned hot stars coincides with the loss of a convective envelope at some critical ``Kraft temperature". We highlighted that this Kraft temperature is expected to decrease with increasing metallicity  (Figure~\ref{fig: metals}). We showed that current observations are consistent with this expectation, but the data is still too sparse to provide a robust test. Thus, we call for more spin-orbit misalignment measurements of hot Jupiters orbiting stars within the temperature range from 6000K to 6300K. This range provides the most promising window into a direct verification of the tidal realignment hypothesis.

\section*{Acknowledgements}

We thank Jeremy Goodman, Konstantin Batygin and Sarah Millholland for useful discussions, in addition to Adrian Barker, Gordon Ogilvie and Yufeng Lin for their invaluable perspective on tidal theory. We are also grateful to Louis Amard for providing open access to the SYCLIST database of stellar evolution models, without which this work would have been substantially more difficult. We would also like to thank the anonymous referee, whose input inspired substantial improvements to the manuscript. This research would not have been possible without the generous support of the Heising-Simons Foundation and their 51 Pegasi b Fellowship (CS). 

\bibliographystyle{aasjournal}
\bibliography{main}

\end{document}